\documentclass[a4paper,10pt]{article}
\usepackage{graphicx}
\usepackage{wrapfig}
\usepackage{subcaption}
\usepackage{amsmath, amssymb}
\usepackage[braket, qm]{qcircuit}
\usepackage{physics}
\usepackage{braket}
\usepackage{relsize}
\usepackage[bookmarksnumbered, colorlinks, linkcolor=blue, urlcolor=blue, citecolor=blue]{hyperref}
\usepackage{enumitem}
\setlist[itemize]{topsep=1pt, itemsep=1pt, parsep=2pt, left=3pt}
\setlist[enumerate]{topsep=1pt, itemsep=1pt, parsep=2pt, left=3pt}
\usepackage{lmodern}
\usepackage{titlesec}
\titlespacing{\section}{0pt}{10pt}{0pt}
\titlespacing{\subsection}{0pt}{5pt}{0pt}
\titlespacing*{\paragraph}{0pt}{2pt}{5pt}
\usepackage{changepage}
\usepackage{biblatex}
\usepackage{booktabs}
\usepackage{amsmath}
\usepackage{mathtools} 
\usepackage[per-mode=symbol]{siunitx} 
\usepackage{physics}
\usepackage{fancyhdr}
\usepackage{setspace}
\usepackage{helvet}

\usepackage{geometry}
\usepackage[bottom]{footmisc}
\usepackage[font=footnotesize]{caption}
\begin{document}
\vspace*{-30pt}
\begin{center}
    \Large\textbf{Gradient-Based Inverse Optimization of\\ Atom-Chip Wire Currents for BEC Transport}\\[1ex]
    \normalsize Naoki Shibuya\\
    Supervised by Dr Fedja Oručević
\end{center}

\begin{center}
    \textbf{\Large Abstract}
\end{center}

\begin{adjustwidth}{2cm}{2cm}
    Modulating wire currents to shift a magnetic trap along an atom chip enables smooth contact-free delivery of Bose–Einstein condensates but can deform the confinement profile causing parametric heating and atom loss. We introduce a fast simulation framework based on inverse optimization that, given an initial trap and a predefined trajectory over time, computes a wire current schedule that transports the atoms and restores the trap geometry upon arrival. We assess trap's minimum energy, lateral displacement, confinement profile and an adiabaticity parameter over a \SI{2.4}{mm} trajectory for various transport durations between \SI{2}{s} and \SI{5}{s}, demonstrating the trade-off between speed and adiabaticity.
\end{adjustwidth}

\section{Introduction}

This section outlines the key concepts and challenges motivating robust BEC transport on atom chips.

\paragraph{Bose--Einstein Condensates (BECs).}
In the 1920s, Bose and Einstein predicted that below a critical temperature, bosonic atoms would accumulate in the ground state, forming a Bose--Einstein condensate (BEC)~\cite{bose1924plancks,einstein1925quantentheorie}. This was first observed in 1995 using dilute $^{87}$Rb gas cooled to nanokelvin temperatures~\cite{anderson1995observation}. BECs are highly coherent and spatially localized, enabling applications in quantum sensing when properly prepared and magnetically confined near atom chips.

\begin{wrapfigure}{r}{0.30\textwidth} 
    \vspace{-30pt}
    \centering
    \includegraphics[width=0.29\textwidth]{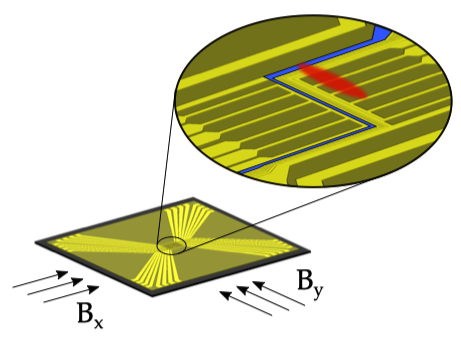}
    \caption{A Z-shaped current path on an atom chip with external bias fields producing an Ioffe-Pritchard trap. Adapted from Barrett~\cite{barrettApparatusProductionBoseEinstein2017}.} 
    \label{fig:barrett-4-33} 
    \vspace{-25pt}
\end{wrapfigure}

\paragraph{Atom Chips and Magnetic Traps.}
Atom chips are compact, flexible platforms for trapping and manipulating ultracold atoms. Microfabricated wires on a planar substrate, combined with external bias fields, generate reconfigurable magnetic potentials that enable precise control of BECs near the chip surface.

\begin{figure}[ht]
    \centering
    \begin{subfigure}[c]{0.28\textwidth}  
        \centering
        \vspace{0pt}
        \includegraphics[width=0.8\textwidth]{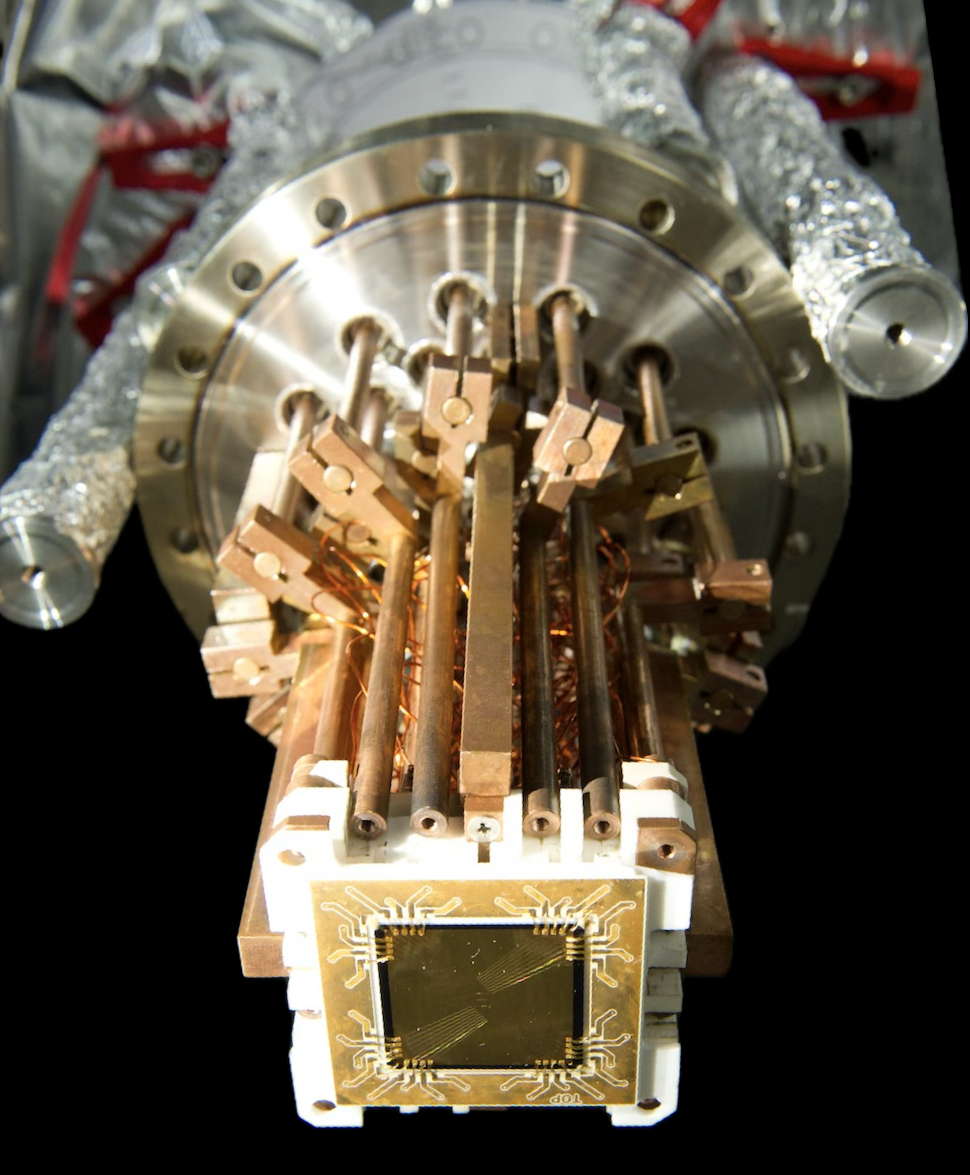}
        \vspace{5pt}
        \caption{Photograph of an atom chip hosted on a printed circuit board (PCB).}
    \end{subfigure}
    \hspace{50pt}  
    \begin{subfigure}[c]{0.38\textwidth}  
        \centering
        \includegraphics[width=0.8\textwidth]{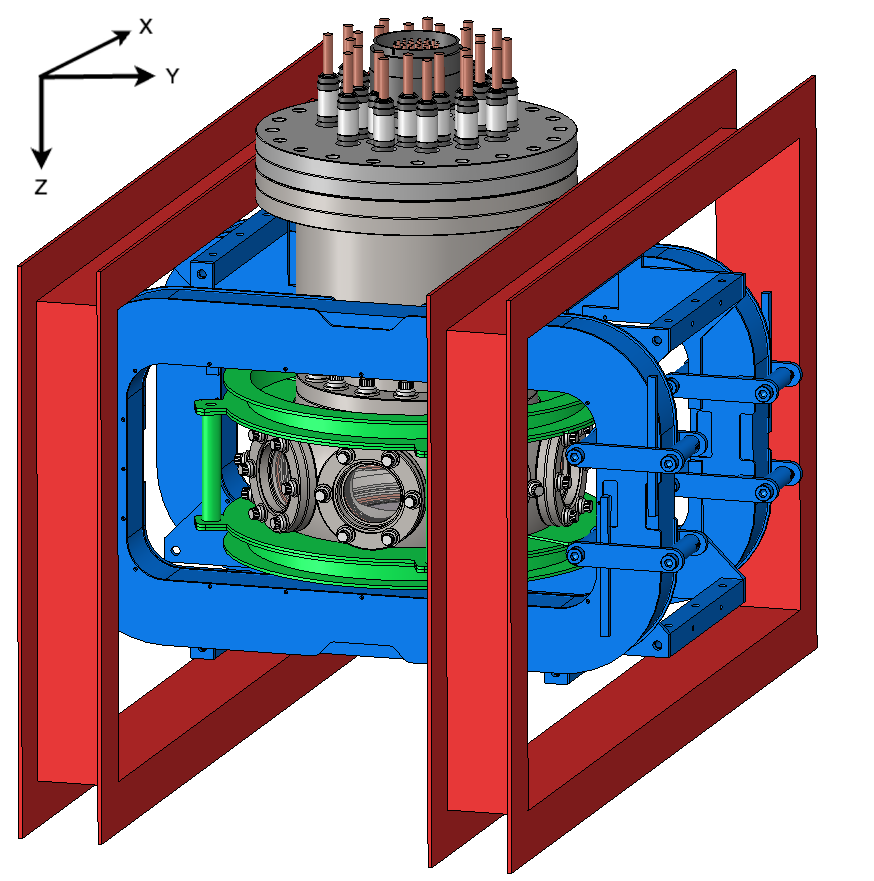}
        \caption{Illustration of a vacuum chamber with anti-reflection-coated windows for optical access. The blue, red, and green frames support external coils for bias fields along the $x$, $y$, and $z$ directions, respectively.}
    \end{subfigure}
    \vspace{-3pt}
    \caption{Adapted from Barrett et al.~\cite{barrettApparatusProductionBoseEinstein2017}.}
    \label{fig:barrett-3-16a-3-20}
\end{figure}

Figure~\ref{fig:barrett-4-33} shows a chip-Z layout, where a Z-shaped wire and external bias fields $B_x$ and $B_y$ form an Ioffe--Pritchard trap with anisotropic confinement and a finite central field~\cite{barrettApparatusProductionBoseEinstein2017}. This suppresses Majorana losses from spin flips in zero-field regions. Figure~\ref{fig:barrett-3-16a-3-20} shows the chip mounted upside down in an ultra-high vacuum (UHV) chamber with optical access for cooling and imaging.

\vspace{-5pt}

\paragraph{BEC Preparation on Atom Chips.}
Neutral $^{87}$Rb atoms are precooled to approximately $200\,\mu\mathrm{K}$ in a mirror magneto-optical trap (MOT) and optically pumped into the low-field-seeking $\ket{F=2,m_F=2}$ Zeeman sublevel~\cite{barrettApparatusProductionBoseEinstein2017}. In this state, the Zeeman shift is given by
\begin{equation}
  \Delta E = g_F\,\mu_B\,m_F\,|B|,
\end{equation}
resulting in a magnetic potential proportional to the local field magnitude $|B|$. The atoms are then transferred into an Ioffe–Pritchard trap on the chip for tighter confinement. Radio-frequency evaporative cooling is applied to reach nanokelvin temperatures, driving the gas into the Bose–Einstein condensed phase. The condensate remains magnetically trapped near the chip surface, ready for transport and interrogation. This state serves as the initial condition for our transport simulations.

\begin{figure}[ht]
    \centering
    \begin{subfigure}[c]{0.45\textwidth}  
        \centering
        \includegraphics[width=0.8\textwidth]{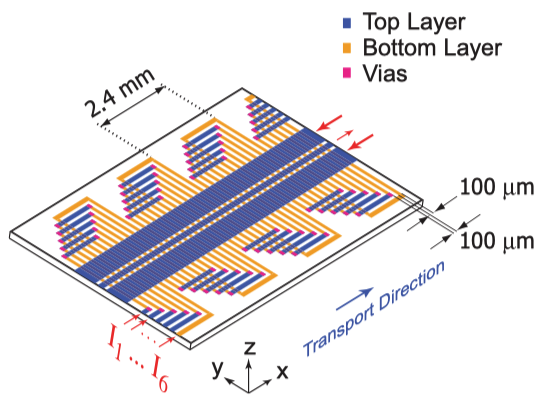}
        \caption{Multilayer atom chip setup}
    \end{subfigure}
    \hspace{30pt}  
    \begin{subfigure}[c]{0.45\textwidth}  
        \centering
        \includegraphics[width=0.9\textwidth]{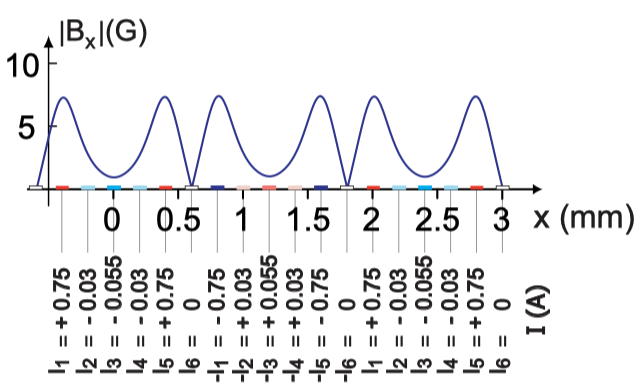}
        \vspace{0pt}
        \caption{Transport potential along the longitudinal direction. The rectangles on the x-axis represent the shifting wires.}
    \end{subfigure}
    \caption{Adapted from Long et al.~\cite{long_2005_LongDistanceMagnetic}}
    \label{fig:long-2-4}
\end{figure}

\paragraph{BEC Transport on Atom Chips.}
Transporting a BEC along an atom chip requires dynamically shifting the magnetic trap minimum without inducing excitations.  A common “magnetic conveyor belt” uses parallel guiding wires to form a static quadrupole trap for lateral confinement and a series of perpendicular shifting wires to create movable Ioffe–Pritchard–like wells as shown in Figure~\ref{fig:long-2-4}. By modulating the shifting-wire currents over time, the trap minimum can be continuously translated along the chip surface.

\begin{wrapfigure}{r}{0.45\textwidth}
    \vspace{-10pt}
    \centering
    \includegraphics[width=0.44\textwidth]{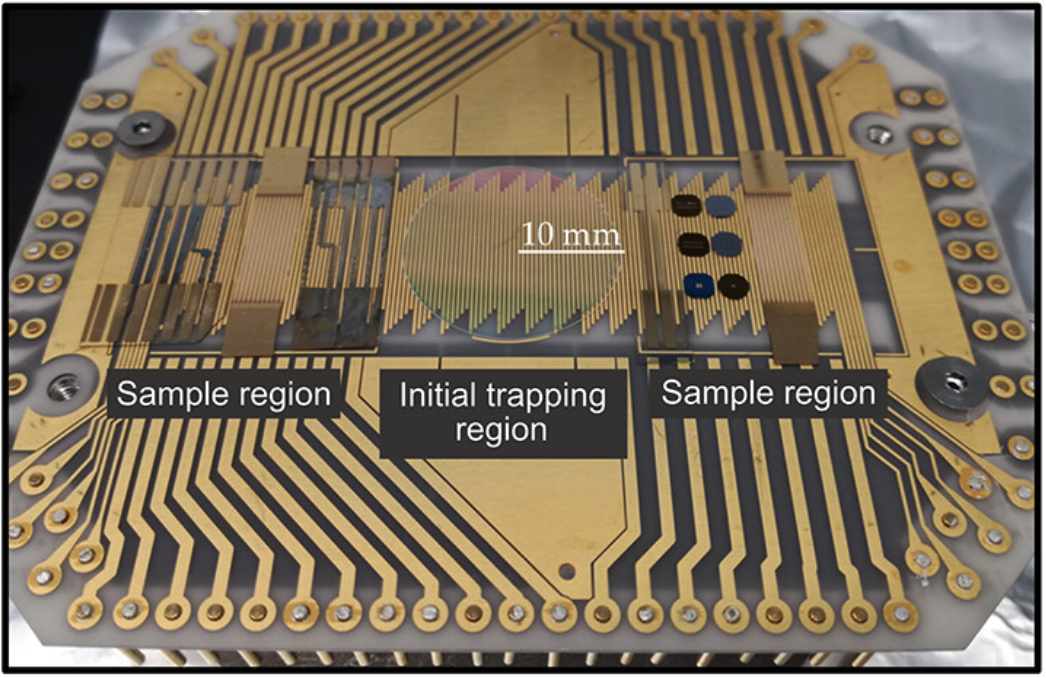}
    \caption{Photograph of a multi-region atom chip. Adapted from Fekete et al.~\cite{feketeQuantumGasEnabledDirect2024}.}
    \label{fig:fekete-5a}
    \vspace{-20pt}
\end{wrapfigure}
Figure~\ref{fig:fekete-5a} shows a more recent multi-region chip design where a central preparation zone hands off condensates to separate imaging regions~\cite{feketeQuantumGasEnabledDirect2024}.

Both architectures face the same core challenge: finding current schedules that preserve trap geometry and ensure adiabatic transport. Even small current imbalances can distort the trap, causing parametric heating, reduced atom number, and degraded coherence. Experimentally tuning these schedules is slow and noise-sensitive, motivating our gradient-based inverse optimization framework for generating reliable BEC transport trajectories in simulation \footnote{The source code is available at \url{https://github.com/naokishibuya/atom-chip-optimizer}}.

\section{Methodology} \label{sec:methodology}

This section presents our overall approach to simulating BEC transport on atom chips, including magnetic field modeling, trap characterization, and current-schedule optimization.

\subsection{Magnetic Field Modeling} \label{sec:magnetic-field-modeling}

The magnetic field is computed by applying the Biot–Savart law to conductive wires modeled as finite rectangular prisms, following the approach in~\cite{barrettApparatusProductionBoseEinstein2017}. For a conductor centered at the origin and carrying current $I$ along the $x$-axis, the magnetic field $\mathbf{B}(\boldsymbol{r})$ at a point $\boldsymbol{r}$ in space is given by:
\begin{equation}
\mathbf{B}(\boldsymbol{r}) = \frac{\mu_0 I}{4\pi} \int_V \frac{(\boldsymbol{r} - \boldsymbol{r}') \times \hat{\boldsymbol{x}}}{|\boldsymbol{r} - \boldsymbol{r}'|^3} \, dV', \label{eq:biot-savart}
\end{equation}
where $\boldsymbol{r}'$ is the integration variable over the wire volume $V$, $\hat{\boldsymbol{x}}$ is the unit vector in the current direction, and $\mu_0$ is the vacuum permeability. This volume integral reduces to a closed-form sum over the eight corners of the prism’s cross-section. The total field at any position is efficiently computed by superposing the contributions from multiple wires and bias fields.

\subsection{Trap Potential and Metrics} \label{sec:trap-characterization-metrics}

The total potential energy of a low-field-seeking $^{87}$Rb atom in a magnetic field plus gravity is:
\begin{equation}
  U(\boldsymbol{r})
  = m_F\,g_F\,\mu_B\,\lvert\mathbf{B}(\boldsymbol{r})\rvert \;-\; m g z,
\end{equation}
where $g_F=0.5$, $\mu_B = \SI{9.27e-24}{\joule\per\tesla}$, $m=\SI{1.44e-25}{\kilogram}$, and $z$ is measured downward from the chip surface, meaning gravitational potential decreases with increasing $z$. Near a local minimum $\boldsymbol{r}_{\min}$, the potential energy $U(\boldsymbol{r})$ can be approximated harmonically:
\begin{equation}
  U(\boldsymbol{r})
  \approx U(\boldsymbol{r}_{\min})
  \;+\;\tfrac12\,(\boldsymbol{r}-\boldsymbol{r}_{\min})^T
    H\,
    (\boldsymbol{r}-\boldsymbol{r}_{\min}), 
\quad
  H_{ij}
  = \left.\frac{\partial^2 U}{\partial x_i\,\partial x_j}\right|_{\boldsymbol{r}_{\min}}.
\end{equation}
Diagonalizing the Hessian matrix $H$ yields eigenvectors $\hat{e}_i$ and eigenvalues $\lambda_i$, which define the trap's principal axes and curvature:
\vspace{-5pt}
\begin{equation}
  H\,\hat{e}_i = \lambda_i\,\hat{e}_i,
  \quad
  \omega_i = \sqrt{\frac{\lambda_i}{m}},  \quad i=1,2,3,
\end{equation}
where $\{\omega_i\}$ is the trap frequencies in \unit{\radian\per\second}. Also, $\omega_i = 2\pi f_i$ where $f_i$ are the trap frequencies in \unit{\hertz}.

\paragraph{Thomas–Fermi Radius.} 
The Thomas–Fermi approximation includes atom–atom interactions in the effective potential and gives the radius of the condensate along axis $i$ as:
\begin{equation}
  R_{\text{TF},i}
  = \biggl(\frac{2\mu}{m\,\omega_i^2}\biggr)^{\!1/2},
\end{equation}
where $\mu$ is the chemical potential. It can be estimated as:
\begin{equation}
  \mu \approx \frac{\hbar \bar{\omega}}{2}\,\left(15\frac{N a_s}{a_{\text{ho}}}\right)^{2/5}, \quad \bar{\omega} = \sqrt[3]{\omega_1 \omega_2 \omega_3}, \quad a_{\text{ho}} = \sqrt{\hbar/m\bar{\omega}},
\end{equation}
where $\bar{\omega}$ is the geometric mean of the trap frequencies, and $a_{\text{ho}}$ is the corresponding harmonic-oscillator length (the spatial scale of the non-interacting ground state). The $s$-wave scattering length is $a_s = \SI{5.2e-9}{\meter}$ for $^{87}\text{Rb}$. We use $N = 10^5$ atoms, typical for atom-chip experiments, yielding a chemical potential of approximately $\SI{3.32e-30}{\joule}$.

\paragraph{Adiabaticity Parameter.}
To quantify how slowly the trap moves relative to its own oscillation and size, we define the instantaneous adiabaticity parameter:
\begin{equation}
  \varepsilon(t)
  = \frac{\lvert v_x(t)\rvert}{\omega_x\,\sigma_x},
\end{equation}
where $v_x(t) = \frac{d}{dt}r_x(t)$ is the transport velocity along $x$, $\omega_x$ is the trap frequency, and $\sigma_x = \frac{R_{\rm TF,x}}{\sqrt{5}}$ is the root-mean-square (RMS) width of the condensate’s Thomas–Fermi profile along the $x$-axis. Since the trap's principal axes may not align exactly with the lab-frame coordinates, $\omega_x$ and $\sigma_x$ are approximated using the principal axis most closely aligned with the $x$-direction. A transport is considered adiabatic when $\varepsilon(t)\ll1$ for all $t$, meaning the trap shifts much more slowly than atoms oscillate within it.

\subsection{Initial Trap Setup} \label{sec:static-potential}

\begin{figure}[ht]
    \centering
    \includegraphics[width=\textwidth]{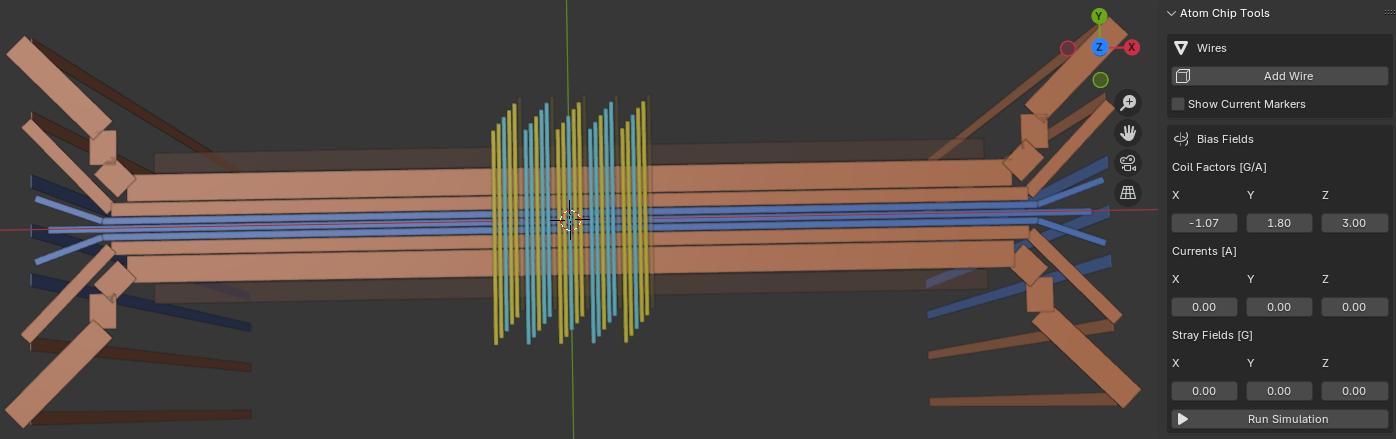}
    \caption{Simulated atom chip layout showing guiding and shifting wires used in BEC transport modeling. Currents alternate between adjacent shifting patterns due to wire continuity. Return paths are omitted because their magnetic contributions are negligible.}
    \label{fig:sim-blender}
\end{figure}

While the actual chip layout is complex, we focus on the wires relevant to transport dynamics: nine guiding wires and five periods of six shifting wires. We ported MATLAB scripts from the University of Sussex to Python/JAX\footnote{JAX: high-performance array computing with automatic differentiation. \url{https://jax.readthedocs.io}} for performance. A Blender plugin\footnote{Blender: 3D creation software. \url{https://www.blender.org}} visualizes JSON-exported wire layouts and supports interactive trap analysis (Figure~\ref{fig:sim-blender}).

\vspace{-5pt}

Each shifting wire is continuous across periods, causing alternating current directions. They carry smaller currents than guiding wires, and their distant return paths contribute negligibly to the magnetic field, so they are excluded from the simulation.

\begin{wrapfigure}{r}{0.47\textwidth}
    \vspace{-15pt}
    \centering
    \includegraphics[width=0.44\textwidth]{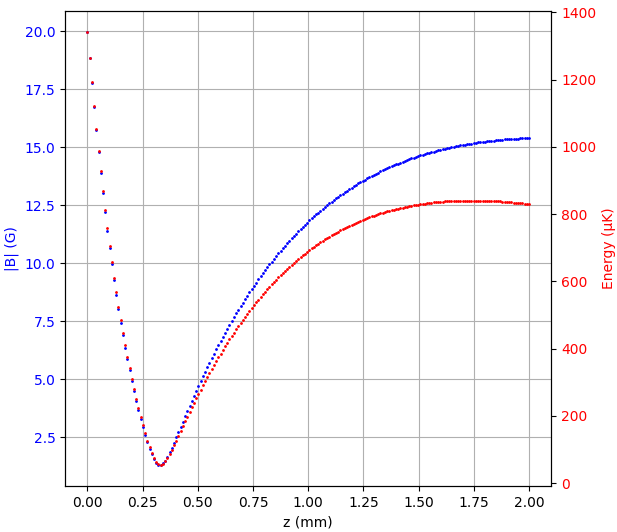}
    \vspace{-5pt}
    \caption{Magnetic field magnitude (blue) and trap potential energy (red) versus the vertical coordinate $z$.}
    \label{fig:sim-potential-1d}
    \vspace{-80pt}
\end{wrapfigure}

At $t = 0$, the six shifting and nine guiding wires have currents:
\vspace{-5pt}
\begin{flalign*}
\mathbf{I}_{\text{shift}}(t=0) &= [\SI{0.6}{\ampere}, \SI{1.05}{\ampere}, \SI{-0.90}{\ampere}, \SI{1.05}{\ampere}, \SI{0.60}{\ampere}, \SI{0.0}{\ampere}] &&\small{\text{left to right}}\\[2pt]
\mathbf{I}_{\text{guide}}(t=0) &= [\SI{0.0}{\ampere}, \SI{13.79}{\ampere}, \SI{13.76}{\ampere}, \SI{-3.78}{\ampere}, \SI{-3.78}{\ampere}, \SI{-3.78}{\ampere}, \SI{13.76}{\ampere}, \SI{13.79}{\ampere}, \SI{0.0}{\ampere}] &&\small{\text{top to bottom}}
\end{flalign*}
\vspace{-5pt}
The Nelder–Mead method\footnote{Implemented using \url{scipy.optimize.minimize}} (robust in non-convex noisy landscapes) finds the central trap minimum as:
\begin{equation*}
  \begin{aligned}
    \boldsymbol{r}_{\min}(t=0) &\approx (0, 0, \SI{0.33}{mm}),\\
    U_{\min}(t=0) &\approx \SI{7.49e-28}{\joule}\; (\SI{54.2}{\micro\kelvin})
  \end{aligned}
\end{equation*}
which is shown in Figure~\ref{fig:sim-potential-1d}, along with the corresponding magnetic field magnitude. 

The eigenvalues and eigenvectors of the Hessian matrix $H(t=0)$ are:
\vspace{-5pt}
\begin{align*}
    \boldsymbol{\lambda}(t=0) &\approx \; [\; 7.1, \;  67.3, \;  74.9 \;] \times \num{e-26}, \\[2pt]
    \hat{\boldsymbol{e}}(t=0) &= \; [\;\hat{e}_1, \;\hat{e}_2, \;\hat{e}_3\;] \\&\approx \begin{bmatrix}
        \begin{array}{r @{.} l r @{.} l r @{.} l}
            0 & 9011 & -0 & 0007 & -0 & 4337 \\
            0 & 4337 &  0 & 0008 &  0 & 9011 \\
           -0 & 0002 & -1 & 0000 &  0 & 0011
        \end{array}
    \end{bmatrix},
\end{align*}
showing that $\hat{e}_1$ and $\hat{e}_3$ are nearly aligned with the $x$-axis and $y$-axis, respectively, while $\hat{e}_2$ points almost directly along the negative $z$-axis.

The corresponding trap frequencies are:
\begin{equation}
    \boldsymbol{\omega}(t=0) = 2\pi \times [\; f_1, \; f_2, \; f_3 \;] \; \approx \; 2\pi \times [\; 111.4, \; 343.8, \; 362.5 \;],
\end{equation}
and the Thomas–Fermi radii are:
\vspace{-5pt}
\begin{equation}
  R_{\text{TF}}(t=0) \approx [\;9.69, \;3.1, \;2.98\;]\;\unit{\micro\meter},
\end{equation}
confirming that the trap is tightly confined transversely and elongated along the longitudinal $x$-direction.

\begin{figure}[ht]
    \centering
    \begin{subfigure}[c]{0.48\textwidth}  
        \centering
        \includegraphics[width=0.9\textwidth]{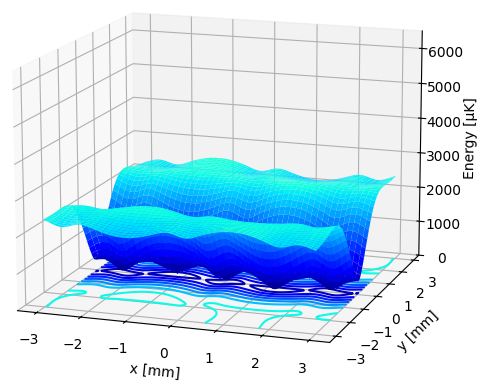}
        \vspace{10pt}
        \caption{3D plot of the potential energy. The chain of potential wells shows longitudinal confinement formed by five periods of the shifting wires.}
    \end{subfigure}
    \hspace{10pt}  
    \begin{subfigure}[c]{0.48\textwidth}  
        \centering
        \includegraphics[width=0.9\textwidth]{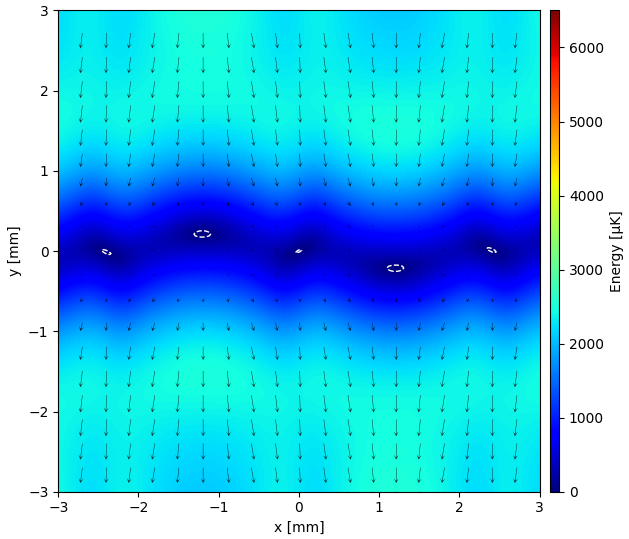}
        \vspace{0pt}
        \caption{2D plot of the potential energy. The dotted lines indicate equipotential lines at $\SI{5}{\micro\kelvin}$ above the minimum. Arrows denote field gradients.}
    \end{subfigure}
    \vspace{0pt}
    \caption{Visualizations of the initial trap landscape.}
    \label{fig:trap-potential}
\end{figure}

Figure~\ref{fig:trap-potential}a shows the potential landscape at the trap minimum height, forming a chain of wells. Figure~\ref{fig:trap-potential}b displays the same potential in the $xy$-plane with magnetic field gradients, clearly revealing five evenly spaced minima along the $x$-axis corresponding to the wire periodicity.

\subsection{Trap Trajectory Scheduling} \label{sec:trap-trajectory-scheduling}

\paragraph{Desired Trap Trajectory.}
Let $\boldsymbol{r}_{\min}(t=0)$ denote the trap center under the initial current configuration. The desired trajectory is defined by:
\begin{equation}
  \boldsymbol{r}_{\text{des}}(t=0) = \boldsymbol{r}_{\min}(t=0),
  \quad
  \boldsymbol{r}_{\text{des}}(t=T) = \boldsymbol{r}_{\min}(t=0) + \Delta x,
\end{equation}
where $T$ is the total duration of the transport process and $\Delta x$ is the total desired displacement.  For intermediate times $t\in[0,T]$, we then define:
\begin{equation}
  \boldsymbol{r}_{\text{des}}(t) = \boldsymbol{r}_{\min}(t=0)\;+\;\operatorname{schedule\_function}\!\left(\tfrac{t}{T}\right)\,\Delta x,
\end{equation}
with a $\operatorname{schedule\_function}\!:[0,1]\to[0,1]$ providing a smooth, monotonic interpolation satisfying the boundary conditions:
$\operatorname{schedule\_function}(t=0)=0$ and $\operatorname{schedule\_function}(1)=1$. This ensures that the trap begins at its initial physical position and follows a smoothly interpolated trajectory to the final destination.

\paragraph{Scheduling Function.}
Our default scheduling function is:
\begin{equation}
    \operatorname{smoothstep}(s) = 10s^3 - 15s^4 + 6s^5, \quad s \in [0, 1],
\end{equation}
which is a quintic polynomial that ensures $C^2$ continuity. We also provide alternative scheduling functions, such as $\operatorname{linear}(s) = s$, which linearly interpolates between 0 and 1, and $\operatorname{cosine}(s) = \frac{1 - \cos(\pi s)}{2}$, which gives a smooth cosine-shaped transition.

We analyze a transport distance of $\Delta x = \SI{2.4}{mm}$ across six shifting wires. Ideally, the trap would shift by $\Delta x$ and recover its original geometry, simply translated along the chip. This would enable seamless and repeatable transport. In practice, our design includes only five periods of shifting wires, so edge effects at both ends break this symmetry. Extending to longer, fully periodic transport cycles would require additional wire segments to suppress boundary distortions. We leave this for future work.

\paragraph{Choice of Step Count.}
To ensure adiabaticity, we first fix the number of discrete steps $N$ such that each spatial increment $\delta x = \Delta x / N$ is much smaller than the initial Thomas–Fermi radius $R_{\text{TF},1}(t=0)$:
\begin{equation}
    \delta x = \frac{\Delta x}{N} \ll R_{\text{TF},1}(t=0)
    \quad\Longrightarrow\quad
    N \gg \frac{\Delta x}{R_{\text{TF},1}(t=0)} = \frac{\SI{2.4}{mm}}{\SI{9.69}{\micro\meter}} \approx 248
\end{equation}
This condition ensures that the trap geometry does not change significantly between steps. As a result, the BEC remains adiabatically trapped throughout the transport, assuming sufficient trap stability. A larger $N$ further reduces per-step deviations, but increases computational cost and may introduce numerical instability. In addition, the choice of scheduling function affects the effective step size, since not all functions yield uniform motion over time. We choose $N = 2500 > 248 \times 10$ to provide a conservative margin.

\paragraph{Transport Time and Speed.}
Once the spatial resolution of the trajectory is set, the remaining condition for adiabaticity depends on the transport speed. At each step $t$, we define the adiabaticity parameter:
\begin{equation}
  \varepsilon(t) = \frac{|v_x(t)|}{\omega_1(t)\,\sigma_1(t)},
  \quad
  v_x(t) = \frac{\delta x(t)}{T},
  \quad
  \sigma_1(t) = \frac{R_{\text{TF},1}(t)}{\sqrt{5}},
\end{equation}
where $\omega_1(t)$ is the trap frequency along the principal axis most closely aligned with the transport direction ($x$-axis), and $\sigma_1(t)$ is the corresponding root-mean-square (RMS) width of the condensate. Ideally, $\varepsilon(t) \ll 1$ is maintained throughout the trajectory to ensure adiabatic transport. After simulating the full path, we evaluate $\varepsilon(t)$ for a range of representative durations such as 2, 3, 4, and 5 seconds.

\subsection{Inverse Optimization of Wire Currents} \label{sec:inverse-optimization}

\paragraph{Problem Formulation.}
At each step $t$, compute the wire current vectors $\mathbf{I}_{\text{wires}}(t) = \bigl[\mathbf{I}_{\text{shift}}(t),\, \mathbf{I}_{\text{guide}}(t)\bigr]$:
\vspace{-5pt}
\begin{align}
  \mathbf{I}_{\text{shift}}(t) &= [\, I_0(t),\; I_1(t),\; I_2(t),\; I_3(t),\; I_4(t),\; I_5(t) \,], \\
  \mathbf{I}_{\text{guide}}(t) &= [\, I_6(t),\; I_7(t),\; I_8(t),\; I_9(t),\; I_{10}(t),\; I_{11}(t),\; I_{12}(t),\; I_{13}(t),\; I_{14}(t) \,].
\end{align}
The objective is to shift the trap center to follow the desired trajectory \(\boldsymbol{r}_{\text{des}}(t)\), such that:
\begin{equation}
\boldsymbol{r}_{\min}\bigl(t+1;\, \mathbf{I}_{\text{wires}}(t)\bigr) \approx \boldsymbol{r}_{\text{des}}(t+1).
\end{equation}
Solving this inverse problem at each step yields a complete current schedule $\{ \mathbf{I}_{\text{wires}}(t) \}_{t=0}^{N-1}$ that transports the trap along the desired path. To prevent excessive wire heating, we constrain the current amplitudes to remain within physical limits: $\pm\SI{3.5}{\ampere}$ for shifting wires (gold) and $\pm\SI{70}{\ampere}$ for guiding wires (copper). These limits may be adjusted depending on hardware capabilities.

\paragraph{Initialization (t = 0).}
The set of wires whose currents are optimized can be configured. After testing several configurations, we chose to optimize only the six shifting wires (0–5), which are responsible for translating the trap along the longitudinal axis. All guiding wires (6–14) are held fixed at their initial current values. We have found that this reduces parameter coupling and provides more stable optimization behavior.

We begin by evaluating the trap using the initial currents $\mathbf{I}_{\text{wires}}(t=0)$. This yields reference metrics characterizing the initial trap: the minimum position $\boldsymbol{r}_{\min}(t=0)$, potential energy $U_{\min}(t=0)$, trap frequencies $\boldsymbol{\omega}(t=0)$, and Thomas–Fermi radii $\mathbf{R}_{\text{TF}}(t=0)$.

\paragraph{Optimization Loop (for each t = 0, ... ,N-1).}
\begin{enumerate}
    \item Compute the desired displacement at $t$: $\delta \boldsymbol{r}(t) = \boldsymbol{r}_{\text{des}}(t+1) - \boldsymbol{r}_{\min}(t)$.
    \item Ensure that the forward displacement along x satisfies $r_{\text{des},x}(t+1) - r_{\min,x}(t) > \epsilon$, where $\epsilon = \SI{1e-6}{mm}$ is a small threshold to prevent backward steps. Otherwise, keep $\mathbf{I}_{\text{wires}}(t)$ unchanged.

    \item Calculate the Hessian $H(t)$ and then evaluate the Jacobian at the trap minimum ($\boldsymbol{r} = \boldsymbol{r}_{\min}(t)$):
    \vspace{-5pt}
    \begin{equation}
      J(t) = \frac{\mathrm{d}\boldsymbol{r}}{\mathrm{d}\mathbf{I}_{\text{wires}}}
    \end{equation}
    The trap minimum satisfies the stationarity condition:
    \begin{equation}
      \nabla U(\boldsymbol{r}, \mathbf{I}_{\text{wires}}) = \frac{\partial U}{\partial \boldsymbol{r}} = 0.
    \end{equation}
    Differentiate with respect to $\mathbf{I}_{\text{wires}}$, noting that the trap position $\boldsymbol{r}$ depends on the wire currents:
    \begin{equation}
    \frac{\mathrm{d}}{\mathrm{d}\mathbf{I}_{\text{wires}}} \left( \frac{\partial U}{\partial \boldsymbol{r}} \right)
    = \frac{\partial^2 U}{\partial \boldsymbol{r} \partial \mathbf{I}_{\text{wires}}} 
    + \frac{\partial^2 U}{\partial \boldsymbol{r}^2} \cdot \frac{\mathrm{d}\boldsymbol{r}}{\mathrm{d}\mathbf{I}_{\text{wires}}} = 0.
    \end{equation}
    Solving for the implicit gradient yields:
    \begin{equation}
    J(t) = \frac{\mathrm{d}\boldsymbol{r}}{\mathrm{d}\mathbf{I}_{\text{wires}}} = -H^{-1}(t) \cdot \frac{\partial^2 U(t)}{\partial \boldsymbol{r} \, \partial \mathbf{I}_{\text{wires}}},
    \end{equation}
    which gives the sensitivity of the trap position to changes in wire currents.
    \item The condition number $\kappa(J(t))$ is monitored to assess numerical stability and to adaptively scale the regularization parameter as:
    \begin{equation}
        \alpha = \lambda \cdot (1 + \kappa(J(t))),
    \end{equation}
    where $\lambda$ is a baseline regularization constant (e.g., $\lambda = 10^{-2}$, which is configurable).
    \item Solve the regularized least-squares problem:
    \begin{equation}
      \min_{\delta \mathbf{I}_{\text{wires}}} \left\| J \, \delta \mathbf{I}_{\text{wires}} - \delta \boldsymbol{r} \right\|^2 + \alpha \left\| \delta \mathbf{I}_{\text{wires}} \right\|^2,
    \end{equation}
    where the goal is to minimize the difference between the predicted trap displacement $J \, \delta \mathbf{I}_{\text{wires}}$ and the target displacement $\delta \boldsymbol{r}$, penalizing large current updates for stability. This is a Tikhonov-regularized least-squares problem (ridge regression in statistics) \cite{hastie2009elements}, and its solution is given by:
    \begin{equation}
      \delta \mathbf{I}_{\text{wires}} = \left( J^\top J + \alpha \mathbf{I} \right)^{-1} J^\top \delta \boldsymbol{r},
    \end{equation}
    where $\mathbf{I}$ is the identity matrix.
    \item Update currents by:
    \begin{equation}
      \mathbf{I}_{\text{wires}}(t+1) = \operatorname{clip}\bigl(\mathbf{I}_{\text{wires}}(t) + M \odot \delta \mathbf{I}_{\text{wires}} \bigr)
    \end{equation}
    where $M$ is a binary mask with ones for optimized wires and zeros for fixed ones, and $\operatorname{clip}$ ensures that the updated currents remain within the physical limits. 
    \item Perform a local minimization (Nelder–Mead) to locate the updated trap center $\boldsymbol{r}_{\min}(t+1)$ and evaluate and record the associated trap properties: $U_{\min}(t+1)$, $\boldsymbol{\omega}(t+1)$, and $\mathbf{R}_{\text{TF}}(t+1)$.
\end{enumerate}

\paragraph{Termination.}
After \(N\) steps, we obtain the full current schedule \(\{\mathbf{I}_{\text{wires}}(t)\}\) that transports the trap along the prescribed trajectory \(\boldsymbol{r}_{\text{des}}(t)\) where $t = 0, \ldots, N-1$.

\section{Results} \label{sec:results}

This section presents the results of the inverse optimization, including the optimized wire-current schedules, the resulting trap trajectory, time evolution of trap geometry, and adiabaticity parameter values.

\subsection{Wire-Current Schedules} \label{sec:results-currents}

\begin{figure}[ht]
  \vspace{-5pt}
  \centering
  \includegraphics[width=0.8\textwidth]{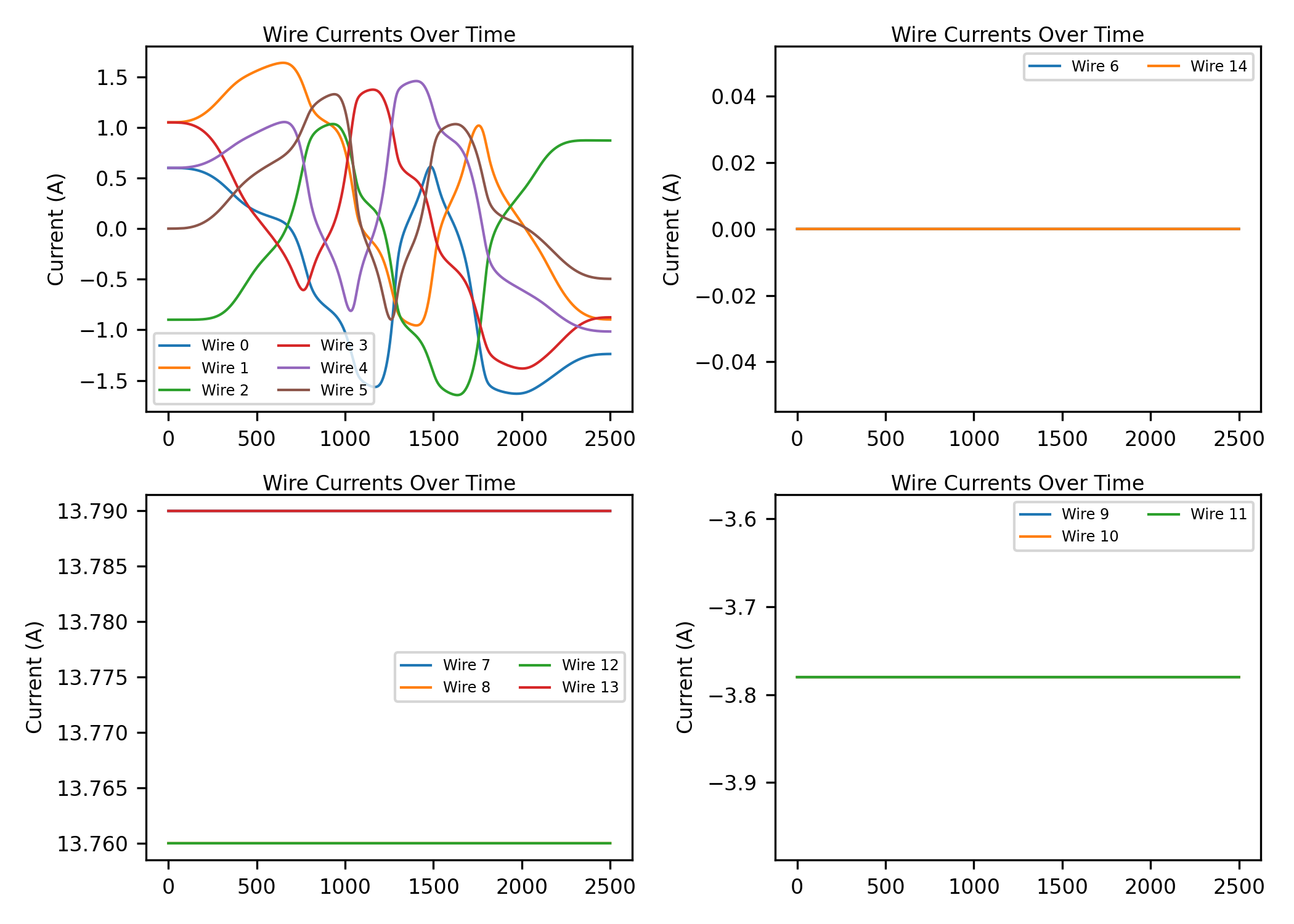}
  \vspace{-5pt}
  \caption{Shifting wires (0–5, top left) are optimized over time, while guiding wires (other panels) remain fixed.}
  \vspace{-5pt}
  \label{fig:currents}
\end{figure}

Figure~\ref{fig:currents} plots the optimized currents for the six shifting wires (wires 0-5) and the nine guiding wires (wires 6-14) over the $N=2500$ steps. 

The shifting-wire currents exhibit small, non-monotonic variations as they incrementally steer the trap minimum toward the target. The guiding-wire currents are fixed at all times, as intended. All currents remain within the predefined limits.

\subsection{Trap Trajectory}

\begin{figure}[ht]
  \centering
  \includegraphics[width=0.99\textwidth]{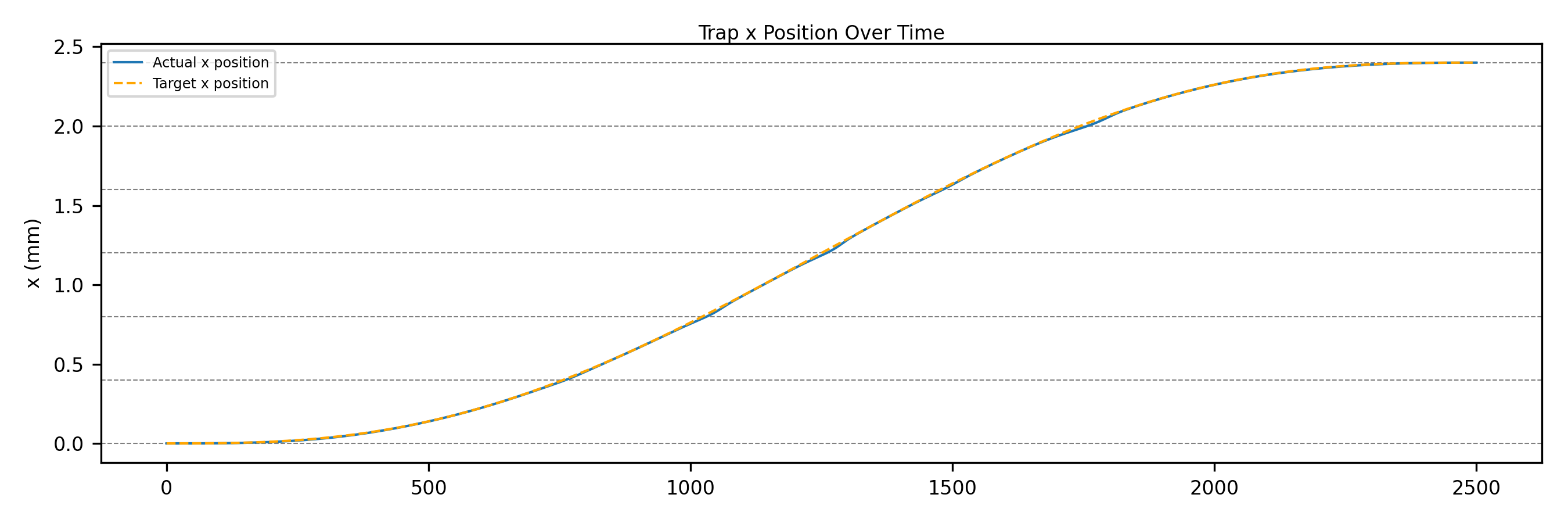}
  \vspace{-10pt}
  \caption{Trap x vs.\ step. The dotted line indicates the center of shifting wires.}
  \label{fig:trajectory}
\end{figure}

\begin{figure}[ht]
  \centering
  \includegraphics[width=0.99\textwidth]{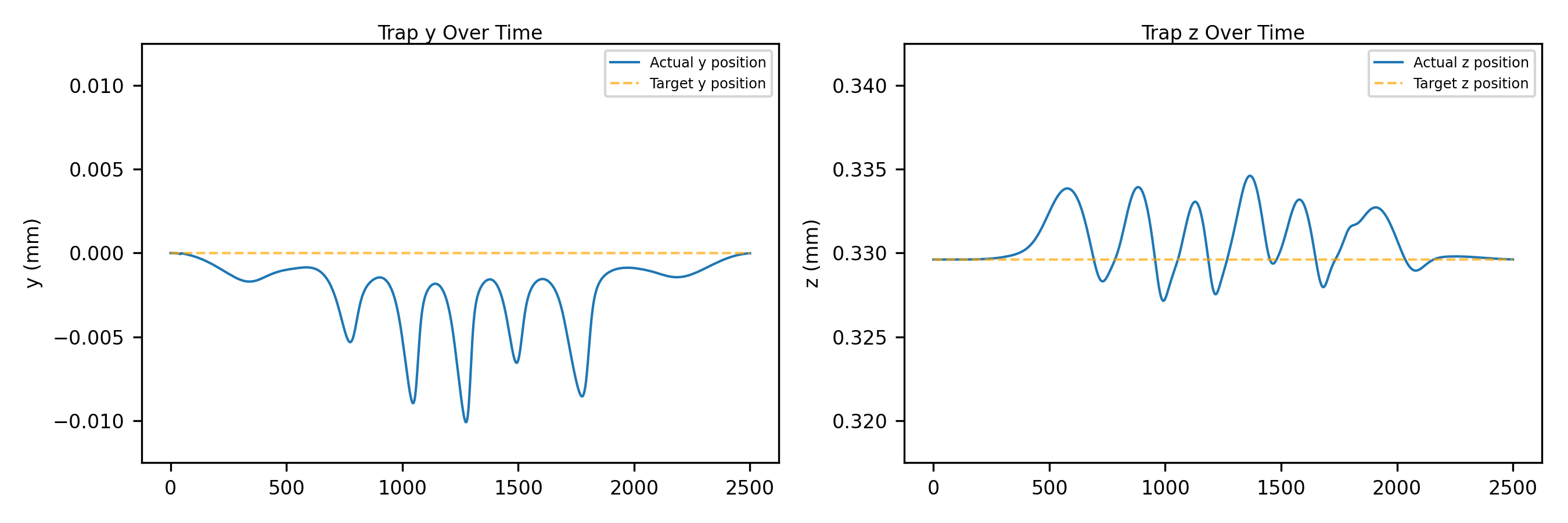}
  \vspace{-10pt}
  \caption{Trap y, z vs.\ step.}
  \label{fig:lateral_position}
\end{figure}

Figure~\ref{fig:trajectory} and Figure~\ref{fig:lateral_position} compare the actual trap center trajectory $\boldsymbol{r}_{\min}(t)$ with the desired path $\boldsymbol{r}_{\text{des}}(t)$. The $x$-component closely follows the smoothstep profile, confirming that the trap progresses as intended along the longitudinal direction.

In the transverse directions, the deviations in $y$ and $z$ remain within $\SI{\pm10}{\micro\meter}$ throughout the transport, indicating excellent accuracy in path tracking across all spatial dimensions.

\begin{figure}[ht]
  \centering
  \includegraphics[width=0.99\textwidth]{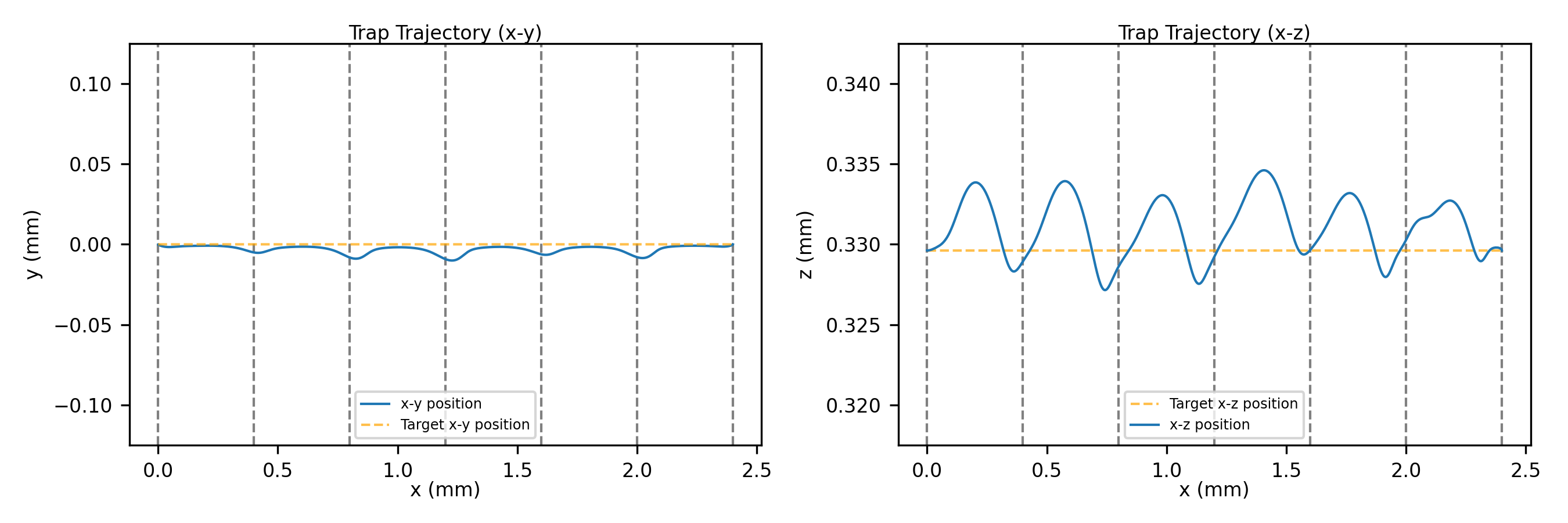}
  \vspace{-10pt}
  \caption{Trap y, z vs.\ x. The dotted line indicates the center of shifting wires.}
  \label{fig:lateral_motion}
\end{figure}

Figure~\ref{fig:lateral_motion} shows the $y$ and $z$ positions versus $x$, revealing periodic lateral motion that arises from the discrete arrangement of the shifting wires.

\subsection{Trap‐Center “Velocity” per Step}
Since the inverse optimization is defined over discrete steps, we report the trap “velocity” in units of $\unit{\micro\meter}/\mathrm{step}$. For each axis $i \in {x, y, z}$ and step $t$, the displacement is:
\begin{equation}
v_i(t) = r_i(t+1) - r_i(t),
\end{equation}
representing the spatial shift of the trap minimum. Given a total duration $T$, dividing $v_i(t)$ by $\delta t = T/N$ yields physical velocity in $\unit{\micro\meter\per\second}$. 

\begin{figure}[ht]
  \centering
  \includegraphics[width=\textwidth]{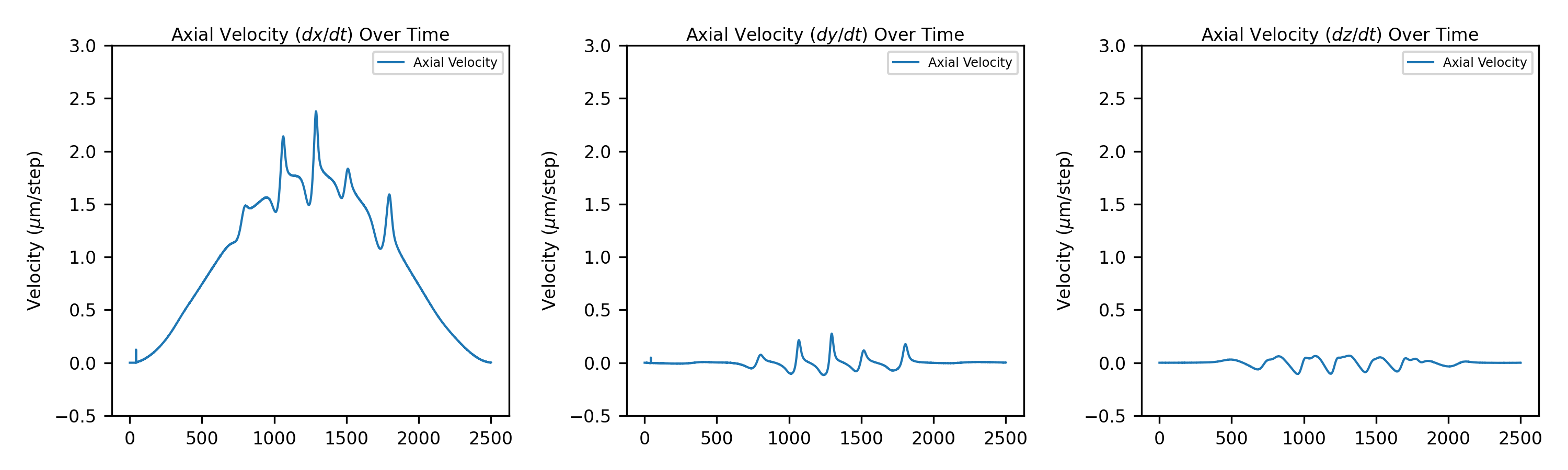}
  \caption{Trap $x$-, $y$-, $z$-velocity per step in $\mu$m/step.}
  \vspace{-10pt}
  \label{fig:velocity}
\end{figure}

Figure~\ref{fig:velocity} shows the stepwise velocities. The $v_x(t)$ qualitatively follows the smoothstep schedule, peaking near the midpoint. The $v_y(t)$ and $v_z(t)$ remain near zero, consistent with the intended axial motion.

The RMS displacement per step are \SI{1.16}{\micro\meter} for $x$, \SI{0.06}{\micro\meter} for $y$, and \SI{0.04}{\micro\meter} for $z$. These are small compared to the initial Thomas–Fermi radii $[\;9.69,\,3.1,\,2.98\;],\unit{\micro\meter}$, confirming that the transport is smooth and spatially bounded, critical for adiabaticity. Minor jitter in $v_x(t)$ appears near the peak but remains well below the trap width.

\begin{figure}[ht]
  \centering
  \includegraphics[width=0.9\textwidth]{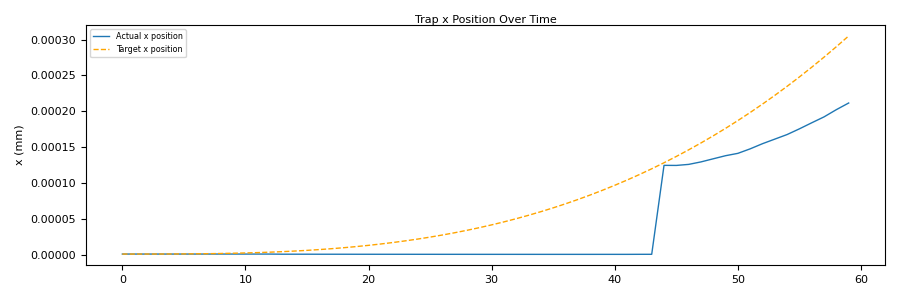}
  \caption{Trap $x$-positions over the first 60 steps.}
  \label{fig:optim_trap_x_60}
\end{figure}

An initial spike in $v_x(t)$ reflects the optimizer’s quick adjustment from rest, necessary due to the initially flat trajectory. This stabilizes rapidly, as shown in Figure~\ref{fig:optim_trap_x_60}.

\subsection{Potential Energies Over Time} \label{sec:results-metrics}

\begin{figure}[ht]
  \centering
  \includegraphics[width=0.8\textwidth]{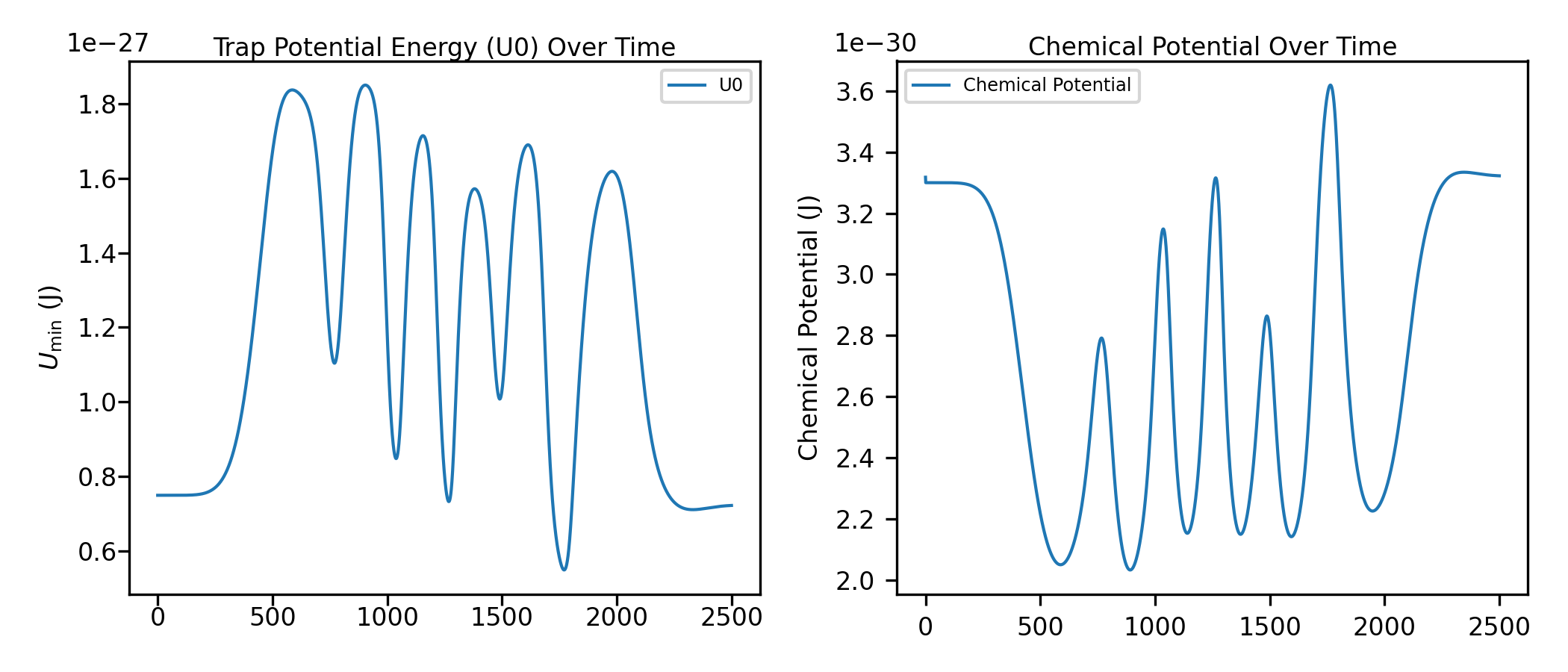}
  \vspace{-10pt}
  \caption{Trap minimum potential $U_{\min}(t)$ (left) and chemical potential $\mu(t)$ (right) over transport steps.}
  \label{fig:optim_potential_chem}
\end{figure}

To evaluate transport stability, we track the trap minimum energy $U_{\min}(t)$ and chemical potential $\mu(t)$ at each step. Figure~\ref{fig:optim_potential_chem} shows that both $U_{\min}(t)$ and $\mu(t)$ fluctuate but at the end, the trap energy remains effectively conserved:
\begin{equation}
    \Delta U_{\min}(t=T) / U_{\min}(t=0) \approx -3.6\%, \qquad
    \Delta\mu(t=T) / \mu(t=0) \approx +0.14\%.
\end{equation}

Per-step energy fluctuations are small relative to the initial values:
\begin{equation}
    \text{RMSE}(\delta U_{\min}) / U_{\min}(t=0) \approx 0.81\%, \quad
    \text{RMSE}(\delta \mu) / \mu(t=0) \approx 0.21\%,
\end{equation}
confirming that the trap energy remains stable throughout the transport.

\subsection{Trap Frequencies and Thomas–Fermi Radii Over Time}

\begin{figure}[ht]
\centering
\includegraphics[width=0.8\textwidth]{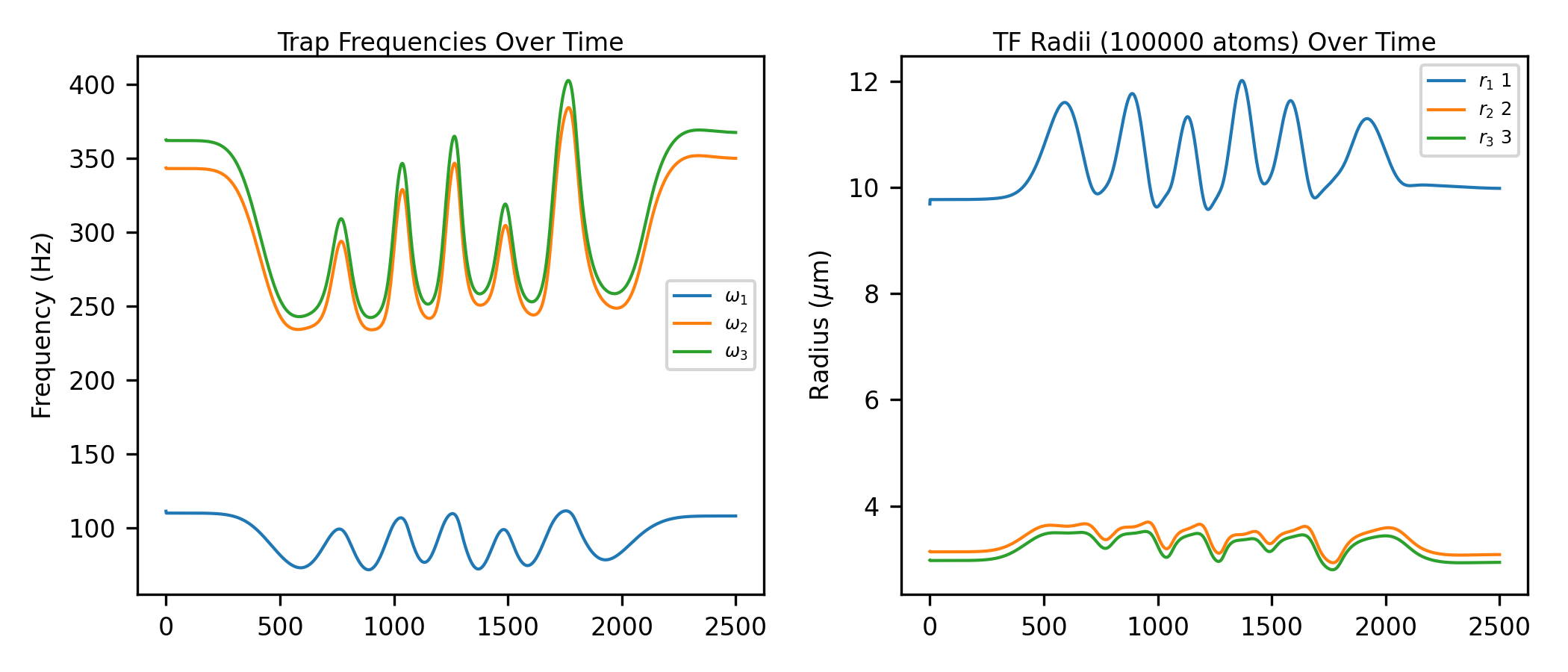}
\caption{Trap frequencies $\omega_{1,2,3}$ and Thomas–Fermi radii $R_{\text{TF},1,2,3}$ over transport steps.}
\label{fig:frequencies-radii}
\end{figure}

Figure~\ref{fig:frequencies-radii} shows the evolution of the trap frequencies $\boldsymbol{\omega}(t)$ and Thomas–Fermi radii $\mathbf{R}_{\text{TF}}(t)$ during transport. The final trap geometry remains close to the initial state:
\begin{equation}
    \Delta\boldsymbol{\omega}(t=T) / \boldsymbol{\omega}(t=0) \approx [\,-2.9\%,\, +1.9\%,\, +1.4\%\,], \qquad
    \Delta\mathbf{R}_{\text{TF}}(t=T) / \mathbf{R}_{\text{TF}}(t=0) \approx [\, +3.0\%,\, -2.0\%,\, -1.3\%\,],
\end{equation}
confirming that the trap frequencies and condensate sizes are well preserved. 

Per-step fluctuations relative to their initial values are:
\begin{equation}
\text{RMSE}(\delta \boldsymbol{\omega}) / \boldsymbol{\omega}(t=0) \approx [\; 19.1\%, \, 19.6\%, \, 20.2\% \;], \quad
\text{RMSE}(\delta \mathbf{R}_{\text{TF}}) / \mathbf{R}_{\text{TF}}(t=0) \approx [\; 9.8\%, \, 9.8\%, \, 10.7\% \;],
\end{equation}
reflecting that the trap geometry oscillates during transport as the condensate crosses the discrete shifting wires.

\subsection{Adiabaticity Verification and Transport Speed} \label{sec:results-trajectory}

\begin{figure}[ht]
  \centering
  \includegraphics[width=0.6\textwidth]{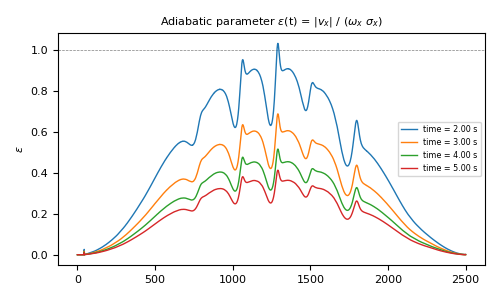}
  \caption{Adiabaticity parameter $\varepsilon(t)$ for $T=2,3,4,5\,\unit{s}$.}
  \label{fig:adiabaticity}
\end{figure}

Figure~\ref{fig:adiabaticity} shows the adiabaticity parameter $\varepsilon(t)=|v_x(t)|/[\omega_1(t),\sigma_1]$ for four total transport durations $T \in \{2,\,3,\,4,\,5\}\,\unit{s}$. For $T = 3$-$5\,\unit{s}$, $\varepsilon(t) < 1$ throughout, confirming adiabatic motion. At $T = 2\,\unit{s}$, $\varepsilon(t)$ briefly exceeds 1 near the midpoint, suggesting the possibility of non-adiabatic excitations.

These results confirm that the optimized current schedule enables smooth, geometry-preserving transport, provided the speed is not too high.

\section{Conclusion}

This thesis presents a numerical framework for transporting a Bose–Einstein condensate (BEC) on an atom chip through inverse optimization of wire currents. By iteratively adjusting control signals based on the gradients of the trap position, the method generates time-dependent current schedules that steer the magnetic trap along a prescribed path while maintaining trap energy, geometry, and adiabaticity.

Future work will incorporate time-dependent Gross–Pitaevskii simulations to assess quantum-state fidelity, extend optimization to two stages, including guiding wires, and explore trajectory prediction using physics-informed machine learning techniques.

The simulation techniques developed here may be extended to other quantum systems requiring precise spatiotemporal control under physical constraints.

\printbibliography

\end{document}